\documentclass[11pt,onecolumn]{IEEEtran}
\usepackage{amsmath}
\usepackage{amssymb}
\usepackage{mathrsfs}
\usepackage{cite}
\usepackage{epsfig}
\usepackage{epsfig}
\usepackage{graphics}
\usepackage{hyperref}
\usepackage{epsfig}

\newtheorem{thm}{Theorem}

\newtheorem{defi}{Definition}

\begin{document}
\title{Secure Joint Source-Channel Coding With Side Information}
\author{Ghadamali Bagherikaram, Konstantinos N. Plataniotis\\
The Edward S. Rogers Sr. Department of ECE, University of Toronto,\\
 10 King's College Road, Toronto, Ontario, Canada M5S 3G4\\
 Emails: \{gbagheri,kostas\}@comm.utoronto.ca
}
 \maketitle
\begin{abstract}
The problem of transmitting an i.i.d Gaussian source over an i.i.d Gaussian wiretap channel with an i.i.d Gaussian side information is considered. The intended receiver is assumed to have a certain minimum SNR and the eavesdropper is assumed to have a strictly lower SNR compared to the intended receiver.  The objective is minimizing the distortion of source reconstruction at the intended receiver.  It is shown, in this work, that unlike the Gaussian wiretap channel without side information, Shannon's source-channel separation coding scheme is not optimum in the sense of achieving the minimum distortion. Three hybrid digital-analog secure joint source channel coding schemes are then proposed which achieve the minimum distortion. The first coding scheme is based on Costa's dirty paper coding scheme and wiretap channel coding scheme when the analog source is not explicitly quantized. The second coding scheme is based on superposition of the secure digital signal and the hybrid digital-analog signal. It is shown that for the problem of communicating a Gaussian source over a Gaussian wiretap channel with side information, there exists an infinite family of optimum secure joint source-channel coding scheme. In the third coding scheme, the quantized signal and the analog error signal are explicitly superimposed. It is shown that this scheme provide an infinite family of optimum secure joint source-channel channel coding schemes with a variable number of binning. Finally, the proposed secure hybrid digital-analog schemes are analyzed under the main channel SNR mismatch. It is proven that the proposed schemes can give a graceful degradation of distortion with SNR under SNR mismatch, i.e., when the actual SNR is larger than the designed SNR.
\end{abstract}
\section{Introduction}
Security protocols are the most critical elements involved in
enabling the growth of the wide range of wireless data networks and
applications. The broadcast nature of wireless communications,
however, makes them particularly vulnerable to eavesdropping. With
the proliferation of more complex modern infrastructure systems,
there is an increasing need for secure communication solutions.
Cryptography is a traditional field that provides
\emph{computationally-secure} protocols at the application layer.
The goal of cryptography has recently been diversified from
providing the critical confidentiality service, to other issues
including authentication, key exchange and management, digital
signature, and more. Unlike the cryptographic approaches, the
recently reintroduced \emph{physical-layer} security aims to develop
effective secure communication schemes exploiting the properties of
the physical layer. This new paradigm can strength the security of
existing systems by introducing a level of
\emph{information-theoretic} security which has provable security,
as compared with computational security. The notion of information
theoretic secrecy in communication systems was first introduced in \cite{1}.
The information theoretic secrecy requires that the received signal
by an eavesdropper not provide any information about the transmitted messages.
Following the pioneering works of \cite{2} and \cite{3} which studied the wiretap
channel, many extensions of wiretap channel model have been considered from a
perfect secrecy point of view (see e.g., \cite{4,5,6,7,8}).

In \cite{9,10}, the Gaussian wiretap channel of \cite{11} is extended to the Gaussian wiretap channel with side information available at the transmitter. In the Gaussian wiretap channel with side information, an i.i.d additive white Gaussian interference is added to the transmitted signal. The interference is completely known at the transmitter an can be used as a covert communication channel. References \cite{9,10} have proposed an achievable perfect secrecy rate for the Gaussian wiretap channel with side information. The achievable coding scheme of \cite{9,10} is a combination of the Costa's dirty-paper-coding \cite{12} and the wiretap channel coding, and the source is assumed to be digital. We refer to this coding scheme as Digital-Secret Dirty-Paper-Coding (DS-DPC) scheme.

All extensions of wiretap channel model have considered communicating a \emph{discrete} source with perfect secrecy constraint. In many applications, however, a bandlimited \emph{analog} source need to be transmitted on a bandlimited Gaussian wiretap channel with side information. In many situations, the exact signal-to-noise ratio (SNR) of the main channel may not be known at the transmitter. Usually, a range of the main channel SNR is known but the true SNR value is unknown. Given a range of main channel SNR such that the eavesdropper's signal is degraded with respect to the  legitimate receiver's signal, it is desirable to design a single transmitter which has a robust performance for all range of SNRs. A common method of designing such a system is based on Shannon's source-channel separation coding: Quantize the analog source and then transmit the result discrete source by the digital secret dirty paper coding scheme. The main advantage of a digital system is that it is more reliable an cost efficient.

The inherent problem of digital systems is that they suffer from a severe form of "threshold effect" \cite{13,14}. This effect can briefly described as follows: The system achieves a certain performance at a certain designed SNR. When the  SNR is increased, the system performance, however, does not improved and it degrades drastically when the true SNR falls below the designed SNR. The severity of the threshold effect in digital systems is related to Shannon's source-channel separation principle \cite{15}. Recent works on non-secure communication systems, however, proved that joint source-channel coding schemes not only can outperform the digital systems for a fixed complexity and delay, they are also more robust against the SNR variations \cite{16,17,18,19,20}.

In \cite{18}, several hybrid digital-analog joint source channel coding scheme is proposed for transmitting a Gaussian source over a (non-secure) Gaussian channel (without side information). The main idea in \cite{18} to increase robustness is to reduce the number of quantization intervals, and thereby increase the distance between the decision lines of the quantization levels. This will, however, increase the distortion. Therefore, to compensate the coarser representation, the quantization error is sent as an analog symbol using a linear coder (see also \cite{21}). In \cite{22}, different coding schemes is analyzed for transmitting a Gaussian source over a Gaussian wiretap channel (without side information). For a fixed information leakage rate to the eavesdropper, \cite{22} has showed that superimposing the secure digital signal with the analog (quantization error) part has better performance compared to the separation based scheme and the uncoded scheme. In \cite{23}, the problem of transmitting a Gaussian source over a (non-secure) Gaussian channel with side information is studied. \cite{23} has introduced several hybrid digital-analog forms of the Costa and Wyner-Ziv coding (\cite{24}) schemes. In \cite{23}, the results of \cite{25} is extended to the case that the transmitter or receiver has side information, and showed that there are infinitely many schemes for achieving the optimal distortion.

In this paper, motivated by a covert communication system, we consider the problem of transmitting an i.i.d Gaussian source over an i.i.d Gaussian wiretap channel with side information available at the transmitter. We assume that the intended receiver has a certain minimum SNR and the eavesdropper has a strictly lower SNR compared to the intended receiver. We are interested in minimizing the distortion of source reconstruction at the intended receiver.  We show that, here, unlike the Gaussian wiretap channel without side information, Shannon's source-channel separation coding scheme is not optimum in the sense of achieving the minimum distortion. We then propose three hybrid digital-analog secure joint source channel coding schemes which achieve the minimum distortion. Our first coding schemes are based on Costa's dirty paper coding scheme and wiretap channel coding scheme when the analog source is not explicitly quantized. We will illustrate that this scheme provide us an extra degree of freedom and therefore we can achieve the optimum distortion. Our second coding scheme is based on superposition of secure digital and hybrid digital-analog signals. We will show that for the problem of communicating a Gaussian source over a Gaussian wiretap channel with side information, there exists an infinite family of optimum secure joint source-channel coding scheme. We explicitly superimpose the quantized and analog signals in our third coding scheme. We will show that this scheme provide an infinite family of optimum secure joint source-channel channel coding schemes with a variable number of binning. Finally, we analyze our secure hybrid digital-analog schemes under the main channel SNR mismatch. We will show that our proposed schemes can give a graceful degradation of distortion with SNR under SNR mismatch, i.e., when the actual SNR is larger than the designed SNR.
\section{Preliminaries And Related Works}
\subsection{Notation}
In this paper, random variables are denoted by capital letters (e.g.
$X$) and their realizations are denoted by corresponding lower case
letters (e.g. $x$). The finite alphabet of a random variable is
denoted by a script letter (e.g. $\mathcal{X}$) and its probability
distribution is denoted by $P(x)$. The vectors will be written as
$x^{n}=(x_{1},x_{2},...,x_{n})$, where subscripted letters denote
the components and superscripted letters denote the vector. The notation
$x_{i}^{j}$ denotes the vector $(x_{i},x_{i+1},...,x_{j})$ for $j\geq i$. A Gaussian
Random variable $X$ with mean of $m$ and variance of $\sigma^{2}$ is denoted by $X\sim \mathcal{N}(m,\sigma^2)$. The function $E[.]$ represents statistical expectation.
\subsection{System Model And Problem Statement}
\underline{Source Model}: Consider a memoryless Gaussian source of $\{V_{i}\}_{i=1}^{\infty}$ with zero mean and variance $\sigma^{2}_{v}$. Thus, $V_{i}\sim \mathcal{N}(0,\sigma^{2}_{v})$ and assume that the sequence $\{V_{i}\}$ is independent and identically distributed (i.i.d). We assume that the source is obtained from uniform sampling of a continuous-time Gaussian process with bandwidth $W_{s}(Hz)$. Furthermore, we assume that the sampling rate is $2W_{s}$ samples per second.

\underline{Channel Model}: The source is transmitted over an Additive White Gaussian (AWGN) wiretap channel in the presence of an interference $S_{i}$ which is known to the transmitter but unknown to the receivers. The channel is modeled as follows:
\begin{IEEEeqnarray}{rl}
Y_{i}=X_{i}+S_{i}+W_{i},\\ \nonumber
Z_{i}=X_{i}+S_{i}+W_{i}^{'},
\end{IEEEeqnarray}
where $X_{i}$, $Y_{i}$ and $Z_{i}$ are the channel input, the received signal by the intended receiver and the received signal by the eavesdropper, respectively. We assume that $E[X_{i}^{2}]\leq P$, and $W_{i}\sim\mathcal{N}(0,N_{1})$, $W_{i}^{'}\sim\mathcal{N}(0,N_{2})$. Furthermore, assume that $S_{i}$'s are a sequence of real i.i.d Gaussian random variables with zero mean and variance $Q$, i.e. $S_{i}\sim \mathcal{N}(0,Q)$. As the source, interference and the channel are i.i.d over the time, we will omit the index $i$ through the rest of the paper. The channel is derived from a continuous-time AWGN wiretap channel with bandwidth $W_{c}(Hz)$. The equivalent discrete-time channel is used at a rate of $2W_{c}$ channel uses per second. The block diagram of the system is depicted in Fig.1.
\begin{figure}[t]
\leftline{\includegraphics[scale=.17]{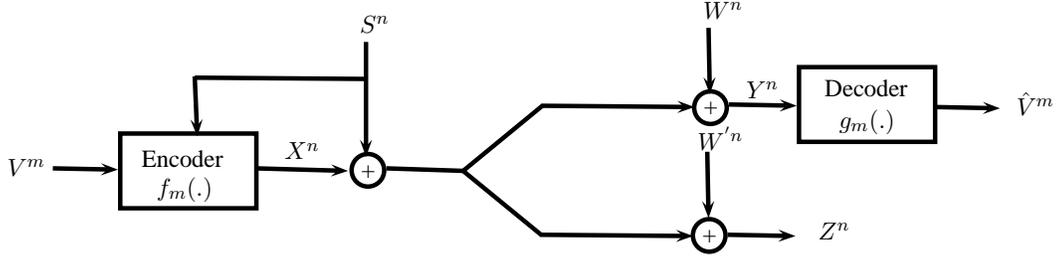}} \caption{Block digram of the secure joint source channel coding problem with Interference known only at the transmitter}
\end{figure}

\underline{Coding Scheme}: The source samples are grouped into blocks os size $m$
\begin{equation}\nonumber
V^{m}=(V_{1},V_{2},...,V_{m}),
\end{equation}
and the encoder is a mapping $f_{m}:\mathbb{R}^{m}\rightarrow \mathbb{R}^{n}$ which satisfies the power constraint $E[\|f_{m}(V^{m})\|^{n}]\leq nP$. Let us define the parameter $\rho=n/m=W_{c}/W_{s}$. In this paper we assume that $\rho=1$. The received signals by the intended receiver and the eavesdropper are given by
\begin{IEEEeqnarray}{rl}
Y^{n}=X^{n}+S^{n}+W^{n},\\ \nonumber
Z^{n}=X^{n}+S^{n}+W^{'n},
\end{IEEEeqnarray}
where $X^{n}=f_{m}(V^{m})$, $W^{n}\sim \mathcal{N}(0,N_{1}\mathbf{I_{n}})$,  $W^{'n}\sim \mathcal{N}(0,N_{2}\mathbf{I_{n}})$, and $\mathbf{I_{n}}$ is the $n\times n$ identity matrix. The decoder at the intended receiver is a mapping $g_{m}:\mathbb{R}^{n}\rightarrow \mathbb{R}^{m}$. The secrecy of the system is measured by the information leaked to the eavesdropper and is expressed as $I_{\epsilon}=\frac{1}{n}I(V^{m};Z^{n})$. Note that $I_{\epsilon}=0$ corresponds to perfect secrecy condition and implies that the eavesdropper obtains no information about the source. In this paper we consider the perfect secrecy situation. The average squared-error distortion of the coding scheme at the intended receiver is given by
\begin{equation}
D_{ave,m}(f_{m},g_{m},N_{1},N_{2})=\frac{1}{m}E[\|V^{m}-\hat{V}^{m}\|^{2}],
\end{equation}
where $\hat{V}^{m}=g_{n}(Y^{n})$. For the purpose of analysis, we will consider sequences of codes $(f_{n},g_{n})$, where $m$ is increasing but the ration $\rho=n/m$ is fixed. The asymptotic performance of the code is given by
\begin{equation}
D_{ave}(N_{1},N_{2})=\lim_{m\rightarrow \infty}D_{ave,m}(f_{m},g_{m},N_{1},N_{2}).
\end{equation}
Note that the above $D_{ave}$ is also a function of $\sigma^{2}_{v}>0$, $P>0$, $Q>0$, and $\rho>0$, but we assume that these parameters are known and fixed, and therefore express $D_{ave}$ as a function of $(N_{1},N_{2})$. In subsequent sections, we refer to $D_{ave}$ as mean-squared distortion and omit the superscript "$ave$" and denote it by $D$, i.e., $D=D_{ave}$.

\underline{Distortion Exponent}: In practical scenarios, the transmitter usually does not have an exact knowledge of $N_{1}$ but knows that $N_{1}\leq N_{d}$, where $N_{d}$ is the noise variance corresponding to the design $SNR_{d}=\frac{P}{N_{d}}$. The eavesdropper channel is still a degraded version of the main channel and is assumed to have the lowest $SNR_{2}<SNR_{d}<SNR_{1}$, where $SNR_{2}=\frac{P}{N_{2}}$ and $SNR_{1}=\frac{P}{N_{1}}$. The receiver is assumed to have a perfect estimate of $SNR_{1}$, but the transmitter communicates at a lower designed $SNR_{d}$. In this scenario, we expect a graceful degradation of distortion $D(SNR_{1})$ with $SNR_{1}$ compared with $D(SNR_{d})$ when the actual $SNR_{1}>SNR_{d}$. Let define the distortion exponent as follows:
\begin{defi}
For a fixed $SNR_{2}$, the distortion exponent of $D(SNR_{1})$ is given by
\begin{equation}\label{eqde}
\zeta\stackrel{\triangle}{=}-\lim_{SNR_{1}\rightarrow \infty}\frac{\log D(SNR_{1})}{\log SNR_{1}}.
\end{equation}
\end{defi}
The highest possible distortion exponent is $\rho$ and therefore, $0\leq \zeta\leq \rho$. The distortion exponent can be used as a criterion for the robustness of a coding scheme. A high distortion exponent means that the coding scheme is more robust. In this paper, we propose different robust coding schemes which achieve the optimum mean-squared distortion. Before introducing our proposed schemes, we need to review some related works in this area.
\subsection{Related Works}
\subsubsection{Digital Wiretap Channel}
In a digital wiretap channel (without any interference $S$), a digital message $M\in\{1,2,...,nC_{s}\}$ is transmitted to the intended receiver while the eavesdropper is kept ignorant. Wyner in \cite{2} characterized the secrecy capacity of this channel when the eavesdropper's channel is degraded with respect to the main channel. Csiszar et. al. in \cite{3} considered the general wiretap channel and established its secrecy capacity. Let us assume $X$, $Y$ and $Z$ be the channel input, intended receiver's signal and eavesdropper's signal, respectively. The secrecy capacity of a wiretap channel is given by
\begin{equation}
C_{s}=2W_{c}\left[I(U;Y)-I(U;Z)\right],
\end{equation}
where $U\rightarrow X\rightarrow YZ$ forms a Markov chain. When the channels are AWGN, \cite{11} showed that the secrecy capacity is given by
\begin{equation}
C_{s}=W_{c}\left[\log(1+\frac{P}{N_{1}})-\log(1+\frac{P}{N_{2}})\right].
\end{equation}
Here, we briefly explain the coding scheme. We Generate $2^{nI(U;Y)}$ Gaussian codewords $U^{n}$ and through them uniformly at random into $2^{nC_{s}}$ bins. Each bin thus contains $2^{nI(U;Z)}$ codeword $U^{n}$. To encode the message $M\in\{1,2,...,2^{C_{s}}\}$ We randomly choose a $U^{n}$ from the bin which is indicated by $M$ and send it. The intended receiver seeks for a $U^{n}$ which is jointly typical with $Y^{n}$ and declare the bin index as the transmitted message. The probability of error asymptotically tends to zero, i.e., $\lim_{n\rightarrow \infty}P_{e}(\hat{M}\neq M)\rightarrow 0$. The information leakage is $\lim_{n\rightarrow \infty}\frac{1}{n}I(M;Z^{n})=0$
\subsubsection{Digital Dirty-Paper Coding}
Consider transmitting a quantized (digital) source over a point-to-point AWGN channel (without any eavesdropper) with side information known at the transmitter. The received signal in this model can therefore be written as $Y=X+S+W$. Costa in \cite{12} showed that the capacity of this channel is given by
\begin{equation}
C=W_{c}\log(1+\frac{P}{N_{1}}),
\end{equation}
which is equal to the capacity of an AWGN channel without the interferer $S$. The achievability scheme- which is called Dirty-Paper Coding (DPC)- is as follows: Let
\begin{equation}
R=2W_{c}\left[I(U;Y)-I(U;S)\right],
\end{equation}
where $U\rightarrow X\rightarrow Y$ forms a Markov chain. We generate $2^{nI(U;Y)}$ Gaussian codewords $U^{n}$ and through them into $2^{nR}$ bins. In each bin there exists $2^{nI(U;S)}$ codewords. To transmit a message $M\in\{1,2,...,2^{nR}\}$, we choose a $U^{n}$ in the bin which is indicated by the message $M$, such that $(U^{n},S^{n})$ are jointly typical. The channel input $X$ is generated as a function of $U$ and $S$ as follows:
\begin{equation}
U=X+\alpha S,
\end{equation}
where $X\sim \mathcal{N}(0,P)$ is independent of $S$ and $\alpha=\frac{P}{P+N_{1}}$. The decoder seeks for a $U^{n}$ which is jointly typical with $Y^{n}$ and declare the bin number as the transmitted message. The channel output is $Y=X+S+W=U+(1-\alpha)S+W$ and it is easy to show that with $\alpha=\frac{P}{P+N_{1}}$, the achievable rate $R=C$.
\subsubsection{Digital Secret Dirty-Paper Coding}
Consider transmitting a digital source over a Gaussian wiretap channel in the presence of the interference $S$ known only at the transmitter. The received signals by the intended user and eavesdropper are given by
\begin{IEEEeqnarray}{lr}
Y=X+S+W,\\ \nonumber
Z=X+S+W^{'},
\end{IEEEeqnarray}
Even in a Gaussian case, the secrecy capacity of this problem is still unknown. We can get an achievable rate by combining the secrecy coding of wiretap channel with Costa's Dirty-Paper-Coding. We refer to this coding scheme as Digital Secret-Dirty-Paper-Coding (DS-DPC).  The following theorem characterizes the best known achievable secure rate for the Gaussian wiretap channel with side information using DS-DPC scheme:
\begin{thm}(\cite{9,10})
For the Gaussian wiretap channel with side information, an achievable secrecy rate is given by
\begin{equation}\label{eq15}
R_{s}=2W_{c}\max_{\alpha}\min\{I(U_{\alpha};Y)-I(U_{\alpha};S), I(U_{\alpha};Y)- I(U_{\alpha};Z)\},
\end{equation}
where $0\leq \alpha \leq 1$ is a real number, $U_{\alpha}= X +\alpha S$, and $X\sim \mathcal{N}(0,P)$.
\end{thm}
The above optimization is further analyzed in \cite{9,10} to provide
the more explicit expression for the secrecy rate. Let
\begin{equation}
R(\alpha)=I(U_{\alpha};Y)-I(U_{\alpha};S),
\end{equation}
and
\begin{equation}
R_{Z}(\alpha)=I(U_{\alpha};Y)-I(U_{\alpha};Z).
\end{equation}
The function $R(\alpha)$ is maximized at $\alpha^{*}=\frac{P}{P+N_{1}}$ and
\begin{equation}
R(\alpha^{*})=\frac{1}{2}\log\left(1+\frac{P}{N_{1}}\right).
\end{equation}
The function $R_{Z}(\alpha)$ is maximized at $\alpha=1$ and
\begin{equation}
R_{Z}(1)=\frac{1}{2}\log\left(\frac{P+Q+N_{1}}{P+Q+N_{2}}\frac{N_{2}}{N_{1}}\right).
\end{equation}
It is easy to show that $R(\alpha_{0})=R_{Z}(\alpha_{0})$, where
\begin{equation}
\alpha_{0}=\frac{PQ+P\sqrt{Q\left(P+Q+N_{2}\right)}}{Q(P+N_{2})}.
\end{equation}
Let
\begin{IEEEeqnarray}{lr}
P_{L}=-N_{1}-\frac{Q}{2}+\frac{\sqrt{Q^{2}+4Q(N_{2}-N_{1})}}{2}\\ \nonumber
P_{H}=-\frac{Q}{2}+\frac{\sqrt{Q^{2}+4QN_{2}}}{2},
\end{IEEEeqnarray}
then the secrecy rate of (\ref{eq15}) can be written as
\begin{equation}\label{eq1}
R_{s}=\left\{
        \begin{array}{ll}
          R(\alpha^{*}), & \hbox{if}~~ P\leq P_{L}; \\
          R(\alpha_{0}), & \hbox{if}~~ P_{L}\leq P \leq P_{H}; \\
          R_{Z}(1), & \hbox{if}~~ P\geq P_{H}.
        \end{array}
      \right.
\end{equation}
The above three cases correspond to the three possible cases of the optimization
problem of $\max_{\alpha}\min\{R(\alpha), R_{Z}(\alpha)\}$: (1) $R(\alpha)$ is optimized at $\alpha= \alpha^{*}$, and $R(\alpha^{*}) < R_{Z} (\alpha^{*})$; (2) $R_{Z} (\alpha)$ is optimized at $\alpha= 1$, and $R(1) > R_{Z} (1)$; and (3) $R(\alpha)$ is
optimized subject to $R(\alpha) = R_{Z} (\alpha)$.

It can be seen from the above equation that the Gaussian wiretap channel with side information has larger secrecy capacity than the Gaussian wiretap channel without the interferer. The side information therefore helps to improve the secrecy capacity. This is in contrast to the point-to-point channel without the secrecy constraint, in which the side information does not affect the capacity.

As showed in \cite{9,10}, the achievable secrecy rate given in (\ref{eq1}) is indeed the secrecy capacity for the cases when $ P\leq P_{L}$ and when $P\geq P_{H}$. The secrecy rate in the former case is the capacity of the channel without secrecy constraint and with the transmitter knowing the state sequence (Dirty-Paper Channel). The secrecy rate in the latter case is the secrecy capacity of an enhanced wire-tap channel, in which the state variable is used as the channel input instead of the channel interference. Both of these two secrecy capacities are clearly upper bounds on the secrecy capacity for the original wiretap channel with side information. Hence, achieving these two bounds imply achieving
the secrecy capacity.
\section{Separation Based Scheme}
Let us consider our problem of transmitting a Gaussian source over the Gaussian wiretap channel with side information as depicted in Fig.1. In this model no leakage information rate is allowed for the eavesdropper. As assumed in this paper $\rho=n/m=1$. In this section, we characterize the achievable distortion when the source $V^{n}$ is first quantized using an optimum quantizer to produce an index $m\in\{1,2,...,2^{nR_{s}}$, where $R_{s}$ is given in (\ref{eq1}). Then, the index $m$ is transmitted using secret dirty paper coding scheme. As the quantizer output is digital information, we refer to this scheme as digital secret dirty paper coding. We briefly review this scheme here. Let $U_{\alpha}=X+\alpha S$, we first generate $2^{nI(U_{\alpha};Y)}$ i.i.d Gaussian sequences $U_{\alpha}^{n}$ and randomly distribute them into $2^{nR_{s}}$ bins such that each bin contains $2^{n\max\{I(U_{\alpha};S),I(U_{\alpha};Z)\}}$. We index each by $m\in\{1,2,...,2^{nR_{s}}\}$. Then place the $2^{n\max\{I(U_{\alpha};S),I(U_{\alpha};Z)\}}$ codewords in each bin randomly into $2^{n\max\{I(U_{\alpha};S),I(U_{\alpha};Z)\}-nI(U_{\alpha};Z)}$ subbins. To send message $m$ with an interference $S^{n}$, the encoder looks in bin $m$ for  sequences $U_{\alpha}^{n}$ such that $(U_{\alpha}^{n},S^{n})$ is jointly typical. The encoder then randomly choose one of the $U_{\alpha}^{n}$ and transmits the associated $X^{n}=U_{\alpha}^{n}-\alpha S^{n}$. The legitimate receiver seeks for a unique sequence $U_{\alpha}^{n}$ such that $(Y^{n},U_{\alpha}^{n})$ is jointly typical and declares the index of the bin containing $U_{\alpha}^{n}$ as the transmitted message.

Since the transmission rate is $R=R_{s}$, the distortion in $V^{n}$ is given by the distortion rate function of $D(R)$. For a Gaussian source and mean-squared error distortion $D(R)=\sigma^{2}_{v}2^{-2R}$ and therefore, the overall distortion is given by
\begin{equation}
D=\sigma^{2}_{v}\times\left\{
    \begin{array}{ll}
      \frac{N_{1}}{P+N_{1}}, & \hbox{if}~~P\leq P_{L}; \\
      \frac{\left(P+\alpha_{0}^{2}Q\right)\left(P+Q+N_{1}\right)-\left(P+\alpha_{0}Q\right)^{2}}{P(P+Q+N_{1})}, & \hbox{if}~~P_{L}\leq P \leq P_{H}; \\
      \frac{P+Q+N_{2}}{P+Q+N_{1}}\frac{N_{1}}{N_{2}}, & \hbox{if}~~ P\geq P_{H}.
    \end{array}
  \right.
\end{equation}
Note that the above distortion value for $P\geq P_{H}$ is still greater than the optimum value, i.e., $D_{opt}=\sigma^{2}_{v}2^{-2I_{\epsilon}}\frac{N_{1}}{P+N_{1}}$. Therefore, while for $P\geq P_{H}$, the channel coding scheme achieves the secrecy capacity, however, the separation abased scheme to transmit a Gaussian source is not optimum. The following theorem illustrates this observation.

\begin{thm}
For transmitting a Gaussian source over a Gaussian wiretap channel with side information knowing at the transmitter, the separation based scheme is not optimum, unless for $P\leq P_{L}$.
\end{thm}

In the following section we show that there exist a few secure joint source channel coding schemes, which are optimum for all range of $P$. Beside achieving the optimum distortion, we will show that our propose secure joint source channel coding are robust when there is an $SNR_{1}$ mismatch. We will analyze the robustness of these scheme later on.
\section{Secure Hybrid Digital-Analog Dirty Paper Coding}
\subsection{Scheme I}
In this section we propose a secure joint source channel coding scheme where the analog source $V^{n}$ is not explicitly quantized. The code construction, encoding and decoding procedures are as follows.

Let us define an auxiliary Gaussian random variable $U_{\alpha,k}$ as follows:
\begin{equation}
U_{\alpha,k}=X+\alpha S+ kV,
\end{equation}
where $X\sim \mathcal{N}(0,P)$, and $X$, $S$ and $V$ are pairwise independent.

\underline{Codebook generation}: We generate $2^{n\left(I(U_{\alpha,k};Y)+I_{\epsilon}\right)}$ i.i.d sequences $U_{\alpha,k}^{n}$, where each component of each sequence is Gaussian with zero mean and variance $P+\alpha^{2}Q+k\sigma^{2}_{v}$. Then, we distribute these codewords into $2^{nR}$ bins, where $R>0$.

\underline{Encoding}: Given the interference sequence $S^{n}$ and the source sequence $V^{n}$, the encoder find  $U_{\alpha,k}^{n}$'s such that $(U_{\alpha,k}^{n},S^{n},V^{n})$ are jointly typical. The encoder then randomly chooses one of them and transmits $X^{n}=U^{n}-\alpha S^{n}-k V^{n}$. If such $U^{n}$'s cannot be found, the encoder declares failure. Let $P_{e1}$ be the probability of an encoder failure.

From the extensions of typicality to infinite alphabet case \cite{26}, we have $\lim_{n\rightarrow \infty}P_{e1}\rightarrow 0$ provided that
\begin{equation}
I(U_{\alpha,k};Y)-R>\max\left\{I(U_{\alpha,k};Z),I(U_{\alpha,k};SV)\right\},
\end{equation}
for any $R>0$. Therefore the following condition must be satisfied:
\begin{equation}\label{eq2}
I(U_{\alpha,k};Y)>\max\left\{I(U_{\alpha,k};Z),I(U_{\alpha,k};SV)\right\}.
\end{equation}

\underline{Decoding}: The legitimate receiver, looks for a unique $U_{\alpha,k}^{n}$ such that $(U_{\alpha,k}^{n},Y^{n})$ is jointly typical and declare $U_{\alpha,k}^{n}$ as the decoder output. It is easy to show that if the condition of (\ref{eq2}) is satisfied the probability that the decoder output is not equal to the encoded $U_{\alpha,k}^{n}$ goes to zero when $n\rightarrow \infty$. The legitimate receiver then estimate the source $V^{n}$ by using the pair $(U_{\alpha,k}^{n},Y^{n})$ as follows:
\begin{equation}
\hat{V}^{n}=\lambda_{1}Y^{n}+\lambda_{2}U_{\alpha,k}^{n},
\end{equation}
where $\lambda_{1}$ and $\lambda_{2}$ are such that minimizes the distortion $D=E[\|V-\hat{V}\|^{2}]$. After some algebra we can see that the optimum values for $(\lambda_{1},\lambda_{2})$ are given by
\begin{IEEEeqnarray}{lr}\label{eq8}
\lambda_{1}=\frac{-k\sigma^{2}_{v}(P+\alpha Q)}{k^{2}\sigma^{2}_{v}(P+Q+N_{1})+(1-\alpha)^{2}PQ+N_{1}(P+\alpha^{2}Q)},\\ \label{eq9}
\lambda_{2}=\frac{k\sigma^{2}_{v}(P+Q+N_{1})}{k^{2}\sigma^{2}_{v}(P+Q+N_{1})+(1-\alpha)^{2}PQ+N_{1}(P+\alpha^{2}Q)}.
\end{IEEEeqnarray}
The MMSE estimation therefore leads to the following distortion:
\begin{equation}\label{eq3}
D(\alpha,k)=\frac{\sigma^{2}_{v}}{1+\frac{k^{2}\sigma^{2}_{v}}{\mu}},
\end{equation}
where $\mu$ is given by
\begin{equation}\label{eq5}
\mu=\frac{(1-\alpha)^2PQ+N_{1}(P+\alpha^2 Q)}{P+Q+N_{1}}.
\end{equation}
 We need to find the optimum values of $(\alpha,k)$ which minimizes the distortion $D$ with the constraint of (\ref{eq2}). we choose $(\alpha,k)$ such that $I(U_{\alpha,k};Y)>I(U_{\alpha,k};SV)>I(U_{\alpha,k};Z)$. This constraint guarantees that the eavesdropper cannot decode $U_{\alpha,k}$. Our optimization problem can therefor be written as
\begin{IEEEeqnarray}{lr}
\min_{(\alpha,k)}D(\alpha,k)\\ \nonumber
\hbox{s.t:}~I(U_{\alpha,k};Y)>I(U_{\alpha,k};SV)>I(U_{\alpha,k};Z).
\end{IEEEeqnarray}

After some manipulation, it is easy to see that the valid region of $(\alpha,k)$ is as follows (see Fig.2. for a typical valid region):
\begin{figure}
\centerline{\includegraphics[scale=.57]{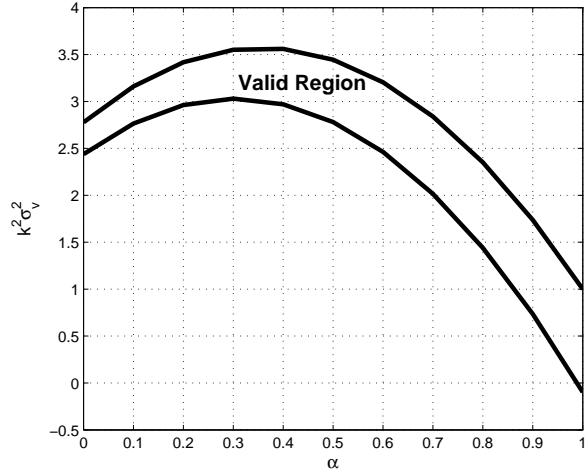}} \caption{A typical valid region for choosing $(\alpha,k)$ in scheme I}
\end{figure}
\begin{IEEEeqnarray}{lr}
\left\{
  \begin{array}{ll}
   k^{2}\sigma^{2}_{v}(P+Q+N_{2})>-Q(P+N_{2})\alpha^{2}+2PQ\alpha+P^{2},\\ \nonumber
k^{2}\sigma^{2}_{v}(P+Q+N_{1})<-(1-\alpha)^{2}PQ-N_{1}(P+\alpha^{2}Q)+P(P+Q+N_{1})
\end{array}
\right.
\end{IEEEeqnarray}
In this region the optimum values of $(\alpha,k)$ and the related $D(\alpha,k)$ are given by
\begin{IEEEeqnarray}{lr}\label{eq10}
\alpha=\frac{P}{P+N_{1}},\\ \label{eq11}
k^{2}=\frac{P^2}{\sigma^{2}_{v}(P+N_{1})}\\
D=\frac{\sigma^{2}_{v}}{\frac{P+N_{1}}{N_{1}}}.
\end{IEEEeqnarray}

The achievable distortion $D$ is indeed optimum and is equal to the distortion of a Gaussian AWGN with side information and without the secrecy constraint. Note that in this scheme, unlike the separation based scheme, we can choose the above optimum $(\alpha,k)$ for all range of $P$. The parameter $K$ (which is due to the analog source), here, provides an extra degree of freedom. Therefore, we can choose the optimum values of $(\alpha,k)$ such that the constraint of $I(U_{\alpha,k};Y)>I(U_{\alpha,k};SV)>I(U_{\alpha,k};Z)$ be satisfied for all $P$.

Note that the proposed secure hybrid digital-analog dirty paper coding is not entirely analog in the sense that the auxiliary random variable $U_{\alpha,k}$ is from a discrete codebook. In contrast to secure digital dirty paper coding however the source is not explicitly  quantized and is embedded into the transmitted signal $X$ in an analog method. An other feature of the secure analog dirty paper coding is that there is no need to double binning the codewords. In the digital scheme however double binning is necessary which one binning is for dirty paper coding and the other binning is for secrecy.

\subsection{Scheme II}
In this scheme, we choose the transmitted signal $X^{n}$ as a superposition  of two signals $X_{1}^{n}$ and $X_{h}^{n}$, which are the outputs of secure digital and hybrid digital-analog dirty paper encoders, respectively. Fig.3. illustrates this scheme.
\begin{figure}
\leftline{\includegraphics[scale=.17]{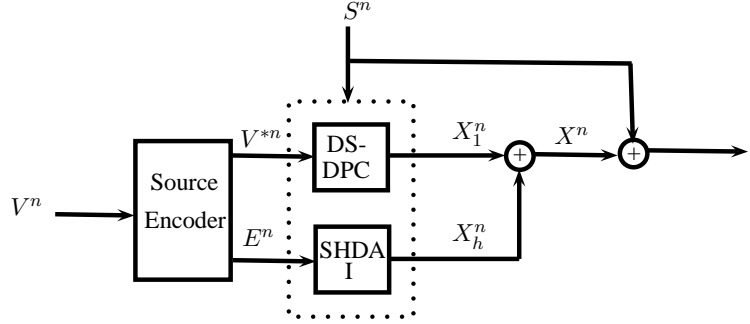}} \caption{Encoder model for secure superposition coding, where DS-DPC is discrete secure dirty paper coding and SHDA I is the hybrid scheme I}
\end{figure}

We first quantize the source at a rate $R<R_{s}$, where $R_{s}$ is given in (\ref{eq1}). Let the quantizer error be $E^{n}=V^{n}-V^{*n}$, where $V^{*n}$ is the reconstruction of $V^{n}$. The quantization error has a variance of $\sigma^{2}_{e}=\sigma^{2}_{v}2^{-2R}$. We encode the quantized digital part using a secure digital dirty paper encoder which treats $S^{n}$ as the interference. The output signal of this section is denoted by $X^{n}_{1}$ which has a power of $P_{1}$. The second part of the encoder is same as scheme I, which treats $S^{n}$ and $X^{n}_{1}$ as interference. The output of this part is denoted by $X^{n}_{h}$, which has a power of $P_{h}=P-P_{1}$. The transmitted signal is the superposition of $X^{n}_{1}$ and $X^{n}_{h}$, i.e., $X^{n}=X^{n}_{1}+X^{n}_{h}$.

We choose the auxiliary random variable of the digital encoder as $U_{1}=X_{1}+\alpha_{1}S$, where $X_{1}$ independent of $S$. The power $P_{1}$ and $\alpha_{1}$ of this encoder are chosen as follows:
\begin{IEEEeqnarray}{lr}
P_{1}=(P+N_{1})(1-2^{-2R}),\\ \nonumber
\alpha_{1}=\frac{P_{1}}{P_{1}+P_{h}+N_{1}}.
\end{IEEEeqnarray}
Note that $P_{1}>0$ and the above choice for $\alpha_{1}$ corresponds to treating $X_{h}$ as noise in addition the channel noise $W$.

In the second encoder, we encode the quantization error $E^{n}$ using the coding scheme I. We choose the parameters of this scheme as follows:
\begin{IEEEeqnarray}{rl}
P_{h}&=(P+N_{1})2^{-2R}-N_{1},\\ \nonumber
\alpha_{h}&=\frac{P_{h}}{P_{h}+N_{1}},\\ \nonumber
k^{2}&=\frac{P_{h}^{2}}{\sigma^{2}_{v}(P_{h}+N_{1})}.
\end{IEEEeqnarray}

Note that as $R\leq \frac{1}{2}\log\frac{P+N_{1}}{N_{1}}$, the power $P_{h}$ is always positive. The auxiliary random is chosen as $U_{\alpha,k}=X_{h}+\alpha_{h}(X_{1}+S)+kE$, where $X_{1}+S$ considered as the net interference. Thus, we choose $X_{h}$ independent of $X_{1}$, $S$ and $E$.

The decoder, first decode the quantization index according the decoding rule of the secure digital dirty paper coding scheme. After that it reconstructs $V^{*n}$. Then, it estimate the quantization error $E^{n}$ according the decoding rule of Scheme I. The overall distortion is the distortion in estimating $E^{n}$. According to the analysis of Scheme I, the overall distortion is given by
\begin{IEEEeqnarray}{rl}
D&=\frac{\sigma^{2}_{e}}{1+\frac{(P+N_{1})2^{-2R}-N_{1}}{N_{1}}}\\ \nonumber
&=\frac{\sigma_{v}^{2}}{\frac{P+N_{1}}{N_{1}}}.
\end{IEEEeqnarray}

As we can see, for any coding rate $R$ of the first encoder such that $R<R_{s}$, this scheme is valid and leads to the optimal distortion. Therefore, we have the following theorem.
\begin{thm}
For the problem of transmitting a Gaussian source over a Gaussian wiretap channel with side information knowing at the transmitter, there exists an infinite family of optimum secure joint source channel coding scheme.
\end{thm}
\subsection{Scheme III}
This scheme is a combination of the scheme I and scheme II. In this scheme the quantized signal and the analog part are explicitly superimposed.

First, we quantize the source $V^{n}$ by using an optimal quantizer at rate $R< R_{s}$. Let $V^{*n}$ and $E^{n}$ be the digital reconstruction of $V^{n}$, and the quantization error vector, respectively. According to the Rate-Distortion theorem, the quantization error $E^{n}$ is a i.i.d Gaussian vector.

We next define an auxiliary random variable $U_{\alpha,k}$ as follows:
\begin{equation}\label{eq6}
U_{\alpha,k}=X+\alpha S+kE,
\end{equation}
where, $X\sim \mathcal{N}(0,P)$, $E\sim \mathcal{N}(0,\sigma^{2}_{v}2^{-2R})$, and $X$, $S$ and $E$ are independent of each other. The coding/decoding procedure is as follows:

\underline{Codebook Generation}: We generate $2^{nI(U_{\alpha,k};Y)}$, i.i.d Gaussian sequences where each component has zero mean and variance of $P+\alpha^{2}Q+k^{2}\sigma^{2}_{v}2^{-2R}$. Then we distribute these codewords into $2^{nR}$ bins and this is shared between the encoder and the decoder. The rate $R$ must be chosen such that each bin contains
\begin{equation}
N_{b}=2^{n\max\{I(U_{\alpha,k};SE),I(U_{\alpha,k};Z)\}}
\end{equation}
codewords. We then randomly distribute the $N_{b}$ codewords of each bin into $N_{sb}$ subbins, where $N_{sb}$ is given by
\begin{equation}
N_{sb}=2^{n\max\{I(U_{\alpha,k};SE),I(U_{\alpha,k};Z)\}-nI(U_{\alpha,k};Z)}.
\end{equation}

\underline{Encoding}: Let $m\in\{1,2,...,2^{nR}\}$ be the quantization index corresponding to the quantized vector $V^{*n}$. To send the message $m$ with an interference $S^{n}$, the encoder looks for $U_{\alpha,k}^{n}$ sequences in the bin $m$ such that $(U_{\alpha,k}^{n},S^{n},E^{n})$ are jointly typical with respect to the distribution of (\ref{eq6}). If such $U^{n}_{\alpha,k}$ cannot be found, encoder declares a failure. Let $P_{e1}$ be the probability of encoding failure. The encoder then randomly choose of the $U^{n}_{\alpha,k}$ and transmits $X^{n}=U^{n}_{\alpha,k}-\alpha S^{n}-kE^{n}$.

\underline{Decoding}: The received signal by the legitimate receiver is $Y^{n}=X^{n}+S^{n}+W^{n}$. The legitimate receiver looks for a unique $U^{n}_{\alpha,k}$ such that is jointly typical with $Y^{n}$. If such a unique $U^{n}$ can be found, the decoder declares $U^{n}$ as the decoder output, otherwise, it declares a failure. Let $P_{e2}$ be the probability of the decoding failure. Then, the legitimate receiver estimates $E^{n}$ from $U^{n}$ and $Y^{n}$.

We can see by similar Gelfand-Pinsker coding argument and Wiretap coding argument that if \begin{equation}\label{eq7}
0\leq R\leq I(U_{\alpha,k};Y)-\max\{I(U_{\alpha,k};SE),I(U_{\alpha,k};Z)\},
\end{equation}
then, $P_{e1}$ and $P_{e2}$ goes to zero when $n\rightarrow \infty$.

\underline{Estimation}: The legitimate receiver makes an MMSE estimate of $V^{n}$ from the observations of $[V^{*n}, U^{n}_{\alpha,k}, Y^{n}]$, where
\begin{IEEEeqnarray}{rl}
V&=V^{*}+E,\\ \nonumber
U_{\alpha,k}&=X+\alpha S+kE, \\ \nonumber
Y&=X+S+W.
\end{IEEEeqnarray}
Let $\sigma^{2}_{e}=\sigma^{2}_{v}2^{-2R}$. The optimum linear MMSE estimation is given by
\begin{equation}
\hat{V}=\lambda_{1}V^{*}+\lambda_{2}U_{\alpha,k}+\lambda_{3} Y,
\end{equation}
where $[\lambda_{1},\lambda_{2},\lambda_{3}]$ are chosen such that minimizes the distortion $D=E[\|V-\hat{V}\|^{2}]$.
We choose the parameters $(\alpha,k)$ as follows:
\begin{IEEEeqnarray}{lr}\label{eq14}
\alpha=\frac{P}{P+N_{1}},\\ \nonumber
k^{2}=P\frac{P+N_{1}-N_{1}2^{2R}}{\sigma^{2}_{v}(P+N_{1})}.
\end{IEEEeqnarray}

We can see the the above choice for $(\alpha,k)$ satisfies the condition of (\ref{eq7}). Let $\mbox{\boldmath{$\Lambda$}}$ be the covariance matrix of $(V^{*n}, U^{n}_{\alpha,k}, Y^{n})^{T}$ and let $\mbox{\boldmath{$\Gamma$}}$ be the correlation vector between $V$ and $(V^{*n}, U^{n}_{\alpha,k}, Y^{n})^{T}$. The matrices of $\mbox{\boldmath{$\Lambda$}}$ and $\mbox{\boldmath{$\Gamma$}}$ then are given by
\begin{IEEEeqnarray}{lr}\label{lm}
\mbox{\boldmath{$\Lambda$}}=\left(
                              \begin{array}{ccc}
                                \sigma^{2}_{v}-\sigma^{2}_{e} & 0 & 0 \\
                                0 & P+\alpha^{2}Q+k^{2}\sigma^{2}_{e} & P+\alpha Q \\
                                0 & P+\alpha Q & P+Q+N_{1} \\
                              \end{array}
                            \right)
\end{IEEEeqnarray}
and
\begin{IEEEeqnarray}{lr}\label{gm}
\mbox{\boldmath{$\Gamma$}}=(\sigma^{2}_{v}-\sigma^{2}_{e}, k \sigma^{2}_{e}, 0)^{T}
\end{IEEEeqnarray}
The coefficients of the linear MMSE estimation are given by $(\lambda_{1},\lambda_{2},\lambda_{3})^{T}=\mbox{\boldmath{$\Lambda$}}^{-1}\mbox{\boldmath{$\Gamma$}}$ and the minimum distortion is given by
\begin{IEEEeqnarray}{lr}
D=\sigma^{2}_{v}-\mbox{\boldmath{$\Gamma$}}^{T}\mbox{\boldmath{$\Lambda$}}^{-1}\mbox{\boldmath$\Gamma$}\\ \nonumber
=\frac{\sigma^{2}_{v}}{\frac{P+N_{1}}{N_{1}}}
\end{IEEEeqnarray}

Note that this scheme is an intermediate scheme between the secure digital dirty paper coding wit the maximum possible binning and the hybrid digital-analog scheme I with the minimum possible bining. We can therefore have a family of schemes with varying bins by changing $R$ such that the condition of (\ref{eq7}) is satisfied.
\begin{thm}
For the problem of transmitting a Gaussian source over a Gaussian wiretap channel with side information knowing at the transmitter, there exists an infinite family of optimum secure joint source channel coding scheme with a variable number of binning.
\end{thm}

Note that the scheme III is closely related to the secure digital dirty paper coding and the scheme II. The difference however is that in the scheme III, the transmitted  signal is not a superposition of the two signals as seen in the scheme II.
\section{Performance Analysis Of The Schemes With SNR Mismatch}
In this section, we analyze the performance of the proposed secure joint source channel coding schemes in the presence of $SNR_{1}$ mismatch. Here, we assume that we have designed the schemes to be optimum for a designed channel $SNR_{d}$, but the actual $SNR_{1}$ is such that $SNR_{2}<SNR_{d}<SNR_{1}$. Separation based digital schemes suffer from the threshold effect. When the actual channel $SNR_{1}$ is worse than the designed $SNR_{d}$, the index cannot be decoded and therefore the distortion drastically increases. When the actual channel $SNR_{1}$ is better than the designed $SNR_{d}$, the distortion is limited by the quantization and does not improve. Thus, the distortion exponent of the separation based scheme is $\zeta=0$.  Our proposed secure hybrid digital-analog schemes however offer better performance in the presence of $SNR_{1}$ mismatch.

\subsection{Performance Analysis of Scheme I}
Let consider the proposed scheme I, which is designed for a $SNR_{d}$ while actual signal-to-noise ratio is $SNR_{1}>SNR_{d}>SNR_{2}$. The receiver can estimate the actual noise power $N_{1}$. Since the receiver knows that the system is designed for $SNR_{d}$, the receiver estimates $\hat{V}$ as follows:
\begin{equation}
\hat{V}=\lambda_{1d}Y+\lambda_{2d}U,
\end{equation}
where $\lambda_{1d}$ and $\lambda_{2d}$ are given in (\ref{eq8}) and (\ref{eq9}), respectively when $(\alpha,k)$ are set as follows:
\begin{IEEEeqnarray}{rl}
\alpha_{d}&=\frac{P}{P+N_{d}},\\ \nonumber k_{d}^{2}&=\frac{P^{2}}{\sigma^{2}_{v}(P+N_{d})}.
\end{IEEEeqnarray}
The received signal is $Y=X+S+W$, where $W\sim\mathcal{N}(0,N_{1})$. The actual $U$ is set by the transmitter as follows:
\begin{equation}
U=X+\alpha_{d}S+k_{d}V,
\end{equation}
The actual distortion is therefore given by
\begin{equation}\label{eq12}
D_{a}(SNR_{1})=\frac{\sigma^{2}_{v}\left(QN_{d}^{2}+\left(P(P+Q)+2PN_{d}+N_{d}^{2}\right)N_{1}\right)}{P^{2}(P+Q)+P(P+Q)N_{d}+QN_{d}^{2}+\left(P(2P+Q)+3PN_{d}+N_{d}^{2}\right)N_{1}}.
\end{equation}
A  useful measure for robustness of a single coding scheme is the rate of decay of the distortion as a function of the actual $SNR_{1}$ when $SNR_{1}\rightarrow\infty$ (see the equation (\ref{eqde})). An upper-bound on the achievable $\zeta$ can be obtained by assuming that a genie informs the transmitter of the actual $SNR_{1}$ and the transmitter chooses an optimum encoding scheme based on the actual $SNR_{1}$. The distortion for the genie-aided scheme is $D=\frac{\sigma^{2}_{v}}{1+SNR_{1}}$. Thus, the distortion exponent is $\zeta=1$. Note that in the absence of any side information the distortion exponent of the genie-aided scheme is $\zeta=1$. In the the absence of the eavesdropper also, the distortion exponent is $\zeta=1$. Therefore, for any single encoding scheme $\zeta\leq 1$.

From equation (\ref{eqde}), the distortion exponent is given by
\begin{equation}\label{eq13}
\zeta=-\lim_{SNR_{1}\rightarrow \infty}\frac{\log D_{a}(SNR_{1})}{\log SNR_{1}}.
\end{equation}
\subsubsection{Absence of Interference}
When there is no interference at the transmitter, i.e., $Q=0$. We can see from the  equations (\ref{eq12}) and (\ref{eq13}) that $\zeta=1$. Therefore, scheme I achieves the optimum distortion $D_{opt}$ as well as the optimum distortion exponent $\zeta=1$.
\subsubsection{Presence of Interference}
In the presence of interference, i.e., $Q\neq 0$, we can see that $\zeta=0$. This is because some residual interference exists in the received signal $Y$, and therefore in hight $SNR_{1}$, the distortion exponent is dominated by this residual interference. However, if optimal distortion is not desired at $SNR_{1}$, then we can achieve the optimum distortion exponent of $\zeta=1$ by using a minor modification in the scheme I. Let modify the auxiliary random variable $U$ in the scheme I as follows:
\begin{equation}
U=X+S+k^{'}V.
\end{equation}
In this modified scheme $\alpha$ is chosen to be 1, which is not an optimum choice for $SNR_{d}$. According to the valid region of Fig.2, the value of  $K^{'}$ must be such that
\begin{equation}
\sqrt{\frac{P^{2}+PQ-QN_{2}}{\sigma^{2}_{v}(P+Q+N_{2})}}<k^{'}<\sqrt{\frac{P^{2}+PQ-QN_{1}}{\sigma^{2}_{v}(P+Q+N_{1})}}.
\end{equation}
Hence, $k^{'}$ can be chosen to be arbitrary close to $\sqrt{\frac{P^{2}+PQ-QN_{1}}{\sigma^{2}_{v}(P+Q+N_{1})}}$. Now $X$ is transmitted and the received signal by the legitimate receiver is $Y$. Using an MMSE estimate for $V$, the final distortion for a actual $SNR_{1}$ when the system is designed for $SNR_{d}$ is given by
\begin{equation}
D_{a}(SNR_{1})=\frac{\sigma^{2}_{v}(P+Q)N_{1}}{(P+Q)N_{1}+k^{'2}_{d}\sigma^{2}_{v}(P+Q+N_{1})},
\end{equation}
where, $k^{'}_{d}=\sqrt{\frac{P^{2}+PQ-QN_{d}}{\sigma^{2}_{v}(P+Q+N_{d})}}$.
Hence, for $SNR_{1}\rightarrow\infty$, the distortion exponent is $\zeta=1$. Fig. 4. shows the distortion of different schemes versus $SNR_{1}$. In this figure $P=1$, $Q=2$, $SNR_{d}=20dB$, $N_{2}=1$, $\sigma^{2}_{v}=1$, and quantization rate is $R=1$.
\begin{figure}
\centerline{\includegraphics[scale=.6]{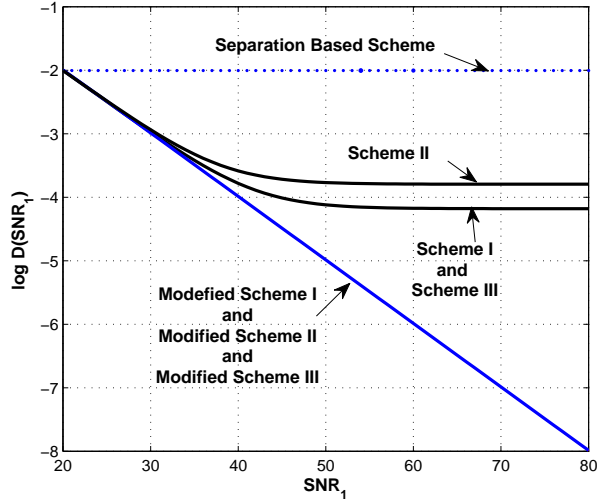}} \caption{Distortion Vs $SNR_{1}$ for different schemes. Here, $P=1$, $Q=2$, $SNR_{d}=20dB$, $N_{2}=1$, $\sigma^{2}_{v}=1$, and $R=1$. }
\end{figure}
\subsection{Performance Analysis of Scheme II}
Let consider the proposed scheme II, in this section. Again, we assume that the scheme is designed for $SNR_{d}$, but the actual is $SNR_{1}>SN_{d}>SNR_{2}$. Scheme II consists of two parts, namely secure digital part and secure hybrid digital-analog part. The performance of the digital part remains constant by increasing the $SNR_{1}$. Thus, performance analysis of this scheme is exactly same as scheme I, when we replace $\sigma^{2}_{v}$ and $P$ with $\sigma^{2}_{e}=\sigma^{2}_{v}2^{-2R}$ and $(P+N_{d})2^{-2R}-N_{d}$, respectively.
\subsection{Performance Analysis of Scheme III}
Next, we analyze the performance of the secure hybrid digital-analog of scheme III under the main channel mismatch.  The different random variables of this scheme are given below.
\begin{IEEEeqnarray}{lr}
U_{d}=X+\alpha_{d}S+k_{d}E,\\ \nonumber
Y=X+S+W,\\ \nonumber
V=V^{*}+E,
\end{IEEEeqnarray}
where $\alpha_{d}=\frac{P}{P+N_{d}}$, $k_{d}^{2}=P\frac{P+N_{d}-N_{d}2^{2R}}{\sigma^{2}_{v}(P+N_{d})}$. The receiver make an MMSE estimate for $V$ as follows:
\begin{equation}
\hat{V}=(\lambda_{1d},\lambda_{2d},\lambda_{3d})(V^{*}, U_{d}, Y)^{T},
\end{equation}
where $(\lambda_{1d},\lambda_{2d},\lambda_{3d})^{T}=\mbox{\boldmath{$\Lambda_{d}$}}^{-1}\mbox{\boldmath{$\Gamma_{d}$}}$, and $\mbox{\boldmath{$\Lambda_{d}$}}$ and $\mbox{\boldmath{$\Gamma_{d}$}}$ are given in (\ref{lm}) and (\ref{gm}), respectively when $(\alpha,k)$ are replaced with $(\alpha_{d},k_{d})$. The actual distortion for the legitimate receiver can therefore written as follows:
\begin{equation}
D_{a}(SNR_{1})=\sigma^{2}_{v}-\mbox{\boldmath{$\Gamma_{d}^{T}$}}\mbox{\boldmath{$\Lambda_{d}^{-1}$}}\mbox{\boldmath{$\Gamma_{d}$}}
\end{equation}
After some math, the actual distortion is given by
\begin{IEEEeqnarray}{rl}
D_{a}(SNR_{1})=\frac{\sigma^{2}_{v}\left(QN_{d}^{2}+\left(P(P+Q)+2PN_{d}+N_{d}^{2}\right)N_{1}\right)}{(P+N_{d})^{2}(P+Q+N_{1})-2^{2R}(N_{d}-N_{1})P(P+Q+N_{d})}.
\end{IEEEeqnarray}
Note that here the actual distortion depends on the quantization rate $R$ and for a special case of $R=0$, the above equation is equivalent to the equation of (\ref{eq12}). Similar to the performance analysis of scheme I, we can see that when $Q=0$, this scheme achieves the optimum distortion exponent of $\zeta=1$. When $Q\neq 0$, the distortion exponent is $\zeta=0$. However, if the optimum distortion at $SNR_{1}$ is not required, we can modify the scheme III by choosing $\alpha=1$ to achieve the optimum distortion exponent $\zeta=1$. As the analysis for this case is straightforward and similar to analysis of scheme I, we therefore omit it here.

\section{Conclusion}
In this work we considered secure joint source channel coding schemes for transmitting an analog source over a Gaussian wiretap channel with known interference at the transmitter. We showed that the separation based coding scheme cannot achieve the minimum distortion. We then proposed a few class of schemes which achieved the optimum distortion. Our proposed schemes are based on Costa's dirty paper coding scheme and Wyner's wiretap channel coding in which the analog source or analog quantization error is not explicitly quantized and embedded in the transmitted signal. We analyzed the performance of the proposed schemes under SNR mismatch. We showed that the proposed schemes can obtain a graceful degradation of distortion with SNR under perfect secrecy constant. The cost of achieving optimum distortion exponent \emph{and} perfect secrecy, is to design the system \emph{near} optimum distortion for a designed SNR. A possible future problem can be if it is still possible to get the optimum distortion exponent under perfect secrecy constraint when we enforce to get optimum distortion for a design SNR. Another possible future work is to design the secure joint source-channel coding schemes for the source-channel bandwidth mismatch ($\rho\neq 1$).


\begin{thebibliography}{9}
\bibitem{1}
C. E. Shannon, ``Communication Theory of Secrecy Systems", {\em Bell
System Technical Journal}, vol. 28, pp. 656-715, October 1949.
\bibitem{2}
A. Wyner, ``The Wire-tap Channel", {\em Bell System Technical
Journal}, vol. 54, pp. 1355-1387, 1975
\bibitem{3}
I. Csisz´ar and J. K¨orner, ``Broadcast Channels with Confidential
Messages", {\em IEEE Trans. Inform. Theory}, vol. 24, no. 3, pp.
339-348, May 1978.

\bibitem{4}
A. Khisti, G. Wornell, A. Wiesel, and Y. Eldar, ``On the Gaussian MIMO Wiretap Channel", {\em in Proc. IEEE International Symposium on Information Theory} , pp.2471-2475, June 2007.
\bibitem{5}
G. Bagherikaram, A. S. Motahari and A. K. Khandani, ``The Secrecy Capacity Region of the Degraded Vector Gaussian Broadcast Channel", {\em in Proc. IEEE International Symposium on Information Theory}, pp.2772-2776, July 2009.
\bibitem{6}
P. K. Gopala, L. Lai and H. El-Gamal, ``On the Secrecy Capacity of Fading Channels", vol. 54, no. 10, pp.4687-4698, October 2008.
\bibitem{7}
Y. Liang; H.V. Poor and S. Shamai, ``Secure Communication over Fading Channels", {\em IEEE Trans. Inf. Theory}, Volume 54, Issue 6 pp: 2470-2492, 2008.
\bibitem{8}
R. Liu, H. V. Poor, ``Secrecy Capacity Region of a Multiple-Antenna Gaussian Broadcast Channel With Confidential Messages", {\em IEEE Trans. Inform. Theory, Volume}, vol. 55, Issue 3, pp.1235-1249, March 2009.


\bibitem{9}
C. Mitrpant, A. J. Han Vinck,  ``An Achievable Region for the Gaussian Wiretap Channel with Side Information ", {\em IEEE Trans. Inf. Theory}, vol. 52, no.5, pp.2181-2190, May 2006.
\bibitem{10}
Y. Chen and H. Vinck, ``Wiretap channel with side information," {\em in Proc. Int. Symp. Inform. Theory}, pp. 2607-2611, Jully 2006.

\bibitem{11}
S. K. Leung-Yan-Cheong and M. E. Hellman, ``Gaussian Wiretap Channel", {\em IEEE Trans. Inform. Theory,} vol. 24, no. 4, pp. 451-456, July 1978.
\bibitem{12}
Max H. M. Costa, ``Writing on dirty paper," {\em IEEE Trans. Info. Theory}, vol. IT-29(3), pp. 439-441, May 1983.
\bibitem{13}
C. E. Shannon, ``Communication in the Presence of Noise ", {\em Proc. IRE}, vol. 39, pp. 10-21, Jan. 1949.
\bibitem{14}
R. J. McAulay and D. J. Sakrison, `` A PPM/PM Hybrid Modulation System", {\em IEEE Trans. Commun. Technol.}, vol. COM-17, pp. 458-469, Aug. 1969.
\bibitem{15}
S. Vembu, S. Verdu, and Y. Steinberg, ``The Source-Channel Separation Theorem Revisited," {\em IEEE Trans. Inform. Theory,} vol. 41, pp. 44-54, Jan. 1995.
\bibitem{16}
Z. Reznic, M. Feder, and R. Zamir, ``Distortion Bounds for Broadcasting With Bandwidth Expansion," {\em IEEE Trans. Inform. Theory,} vol. 52, no. 8,  pp. 3778-3788, Aug. 2006.
\bibitem{17}
B. Chen and G. Wornell, ``Analog error-correcting codes based on chaotic dynamical systems," {\em IEEE Trans. Commun.,} vol. 46, no. 7, pp.
881-890, Jul. 1998.
\bibitem{18}
U. Mittal and N. Phamdo, ``Joint source-channel codes for broadcasting
and robust communication," {\em IEEE Trans. Inf. Theory,} vol. 48, no. 5, pp.
1082-1103, May 2002.
\bibitem{19}
Z. Reznic, R. Zamir, and M. Feder, ``Joint source-channel coding of a Gaussian-mixture source over the Gaussian broadcast channel," {\em IEEE
Trans. Inf. Theory,} vol. 48, no. 3, pp. 776-781, Mar. 2002.
\bibitem{20}
N. Merhav, S. Shamai, ``On joint source-channel coding for the Wyner-Ziv source and the Gel'fand-Pinsker channel,"
{\em IEEE Trans. Inform. Theory,} vol. 49, no. 11, pp. 2844-2855, Nov. 2003.
\bibitem{21}
M. Skoglund, N. Phamdo, and F. Alajaji, ``Design and Performance of vq-Based Hybrid Digital-Analog Joint Source Channel Codes," {\em IEEE Trans. Inform. Theory,} vol 48, no. 3, pp. 708-720, March 2002.
\bibitem{22}
M. P. Wilson, and K. Narayanan, `` Transmitting an Analog Gaussian Source Over a Gaussian Wiretap Channel Under SNR Mismatch," {\em Proc. IEEE International Conference on Telecommunications,} pp.44-47, April 2010.
\bibitem{23}
M. P. Wilson, and K. Narayanan, ``Joint Source Channel Coding with Side Information Using Hybrid Digital Analog Codes," {\em Proc. IEEE Information Theory and Application Workshop}, pp.  299-308, Feb. 2007.
\bibitem{24}
A. D. Wyner and J. Ziv, ``The rate distortion function for source coding with side information at the decoder," {\em IEEE Tran. Info. Theory,} vol. IT-22,
No. 1, pp. 1-10, Jan. 1976.
\bibitem{25}
S. Bross ,A. Lapidoth, and S. Tinguely, ``Superimposed Coded and Uncoded Transmissions of a Gaussian Source over the Gaussian Channel," {\em Proceedings of the IEEE International Symposium on Information
Theory (ISIT),} pp. 2153-2155, July 2006.
\bibitem{26}
T.Cover and J. Thomas, {\em Elements of Information Theory,} Wiley, 2006

\end{thebibliography}
\end{document}